%% file: manuscript.tex
\documentclass[%
 aip,
 amsmath,amssymb,
 reprint,%
]{revtex4-1}

\usepackage{graphicx}
\usepackage{dcolumn}
\usepackage{bm}

\usepackage{graphicx} 
\usepackage{tikz}
\usepackage{tikz-3dplot}
\usepackage{xcolor}
\usepackage{amsmath}
\usepackage[margin=1in]{geometry}
\usepackage{enumitem}
\usepackage[colorlinks=true, linkcolor=blue, citecolor=blue, urlcolor=black]{hyperref}

\usepackage{comment}
\usepackage{float}
\usepackage{xcolor}

\usepackage{booktabs}
\newcounter{fighead}

\makeatletter
\def\@email#1#2{%
 \endgroup
 \patchcmd{\titleblock@produce}
  {\frontmatter@RRAPformat}
  {\frontmatter@RRAPformat{\produce@RRAP{*#1\href{mailto:#2}{#2}}}\frontmatter@RRAPformat}
  {}{}
}%
\makeatother
\begin{document}

\preprint{AIP/123-QED}

\title{Tearing and Kelvin–Helmholtz dynamics in fully kinetic particle-in-cell simulations of electron-scale current sheets}
\author{Sushmita A. Mishra}
\affiliation{Institute of Advanced Research, Koba Institutional Area, Gandhinagar 382 426, India.}
\author{\textsuperscript{*}Gurudatt Gaur}%
 \email{gurudatt.gaur@sxca.edu.in}
\affiliation{St. Xavier's College (Autonomous), Navarangpura, Ahmedabad 380 009, India.}%


\date{\today}

\begin{abstract}
We investigate the stability and nonlinear evolution of localized electron-scale current sheets using fully kinetic, electromagnetic particle-in-cell (PIC) simulations in two and three dimensions. By varying the current-sheet thickness, we examine how it influences the dominant instability and subsequent nonlinear dynamics. In two dimensions, the evolution is governed by electron inertial tearing, with growth rates in good agreement with linear electron magnetohydrodynamics (EMHD) predictions. In three dimensions, however, a thickness-dependent transition emerges. For wider current sheets, a velocity-shear–driven Kelvin–Helmholtz–type instability dominates the early and intermediate evolution, leading to vortex formation and strong modulation of the current layer, followed by the re-emergence of tearing at later times. In contrast, thinner sheets remain tearing-dominated throughout, with no transition to a shear-driven regime, although their effective growth rate is reduced relative to linear predictions, suggesting the influence of mode coupling and three-dimensional effects. These results establish a thickness-dependent transition from tearing-dominated to shear-driven dynamics and reveal a nonlinear sequence of instability evolution in fully kinetic systems, providing new insight into the competition between curvature-driven and shear-driven instabilities in electron-scale current sheets.
\end{abstract}

\maketitle

\section{Introduction}

Current sheets with thickness comparable to the electron skin depth are frequently observed in a wide range of collisionless plasma environments, including the solar wind \cite{phan2018, cheng2018, chen2019}, planetary magnetospheres \cite{burch2016,norgren2018}, astrophysical jets \cite{zenitani2005}, laboratory plasmas \cite{yamada2010, fox2011}, and numerical simulations of plasma turbulence \cite{califano2020} and reconnection \cite{birn2001, karimabadi2013, wan2015}. 
These structures naturally arise through processes such as magnetic reconnection \cite{liu2024,li2023}, turbulent cascade \cite{dong2022,stawarz2024}, and current-driven plasma flows \cite{jain2014,jain2021}. Because of their small spatial scales and strong gradients in magnetic field and electron velocity, electron-scale current sheets are generally unstable and can evolve through a variety of plasma instabilities driven by current gradients, velocity shear, and magnetic curvature.

The evolution of such current sheets is believed to play a crucial role in the conversion of large-scale magnetic energy into particle energy in collisionless plasmas \cite{biskamp,priest,jain2021}. In particular, their instability and nonlinear evolution can lead to rapid magnetic reconnection, particle acceleration, and the conversion of large-scale magnetic energy into heat, even in the absence of classical collisional processes such as viscosity and resistivity \cite{biskamp2006,treumann2001,drake2006}. Understanding the stability and nonlinear evolution of electron-scale current sheets is therefore essential for explaining energy conversion processes in laboratory, space, and astrophysical plasmas.

At electron spatial and temporal scales, the dynamics of such thin current sheets is governed primarily by electron motion, while ions respond only weakly on these timescales and effectively act as a stationary, charge-neutralizing background. Under these conditions, the plasma dynamics can be described within the framework of electron magnetohydrodynamics (EMHD), which is obtained from the two-fluid description by neglecting ion motion while retaining electron inertia in the magnetic-field evolution \cite{kingsep1990,gordeev1994}. EMHD therefore provides a useful reduced model for studying electron-scale current sheets in regimes where the characteristic length scales are comparable to the electron skin depth ($d_e = c/\omega_{pe}$) and the evolution occurs on timescales comparable to $\omega_{pe}^{-1}$, where $\omega_{pe}$ denotes the electron plasma frequency.

Within this framework, several studies have investigated the linear stability and nonlinear evolution of electron-scale current sheets \cite{jain2003, gaur2009}. Electron inertial tearing modes driven by current gradients and magnetic-field curvature have been extensively studied as a mechanism for electron-scale reconnection and magnetic-island formation \cite{bulanov1992,avinash1998,delsarto2003}. In addition to tearing, EMHD systems can support surface-preserving or non-tearing modes associated with magnetic curvature and electron shear flows \cite{lukin2009, mishra2024}. The coexistence and interaction of tearing and surface-preserving modes in two-dimensional EMHD current sheets were examined by Gaur and Kaw \cite{gaur2016}, who showed that equilibrium curvature can allow both instabilities to develop simultaneously and significantly influence the nonlinear evolution of the current layer.

When strong velocity shear is present along the current direction, Kelvin–Helmholtz–type instabilities may also arise in the EMHD regime, leading to vortex formation and deformation of the current sheet \cite{jain2012,gaur2012}. More recently, three-dimensional EMHD studies have demonstrated that the dominant instability in electron current sheets can depend sensitively on the current-sheet thickness and equilibrium shear, with tearing modes and shear-driven Kelvin–Helmholtz–type modes competing in thin electron current layers \cite{jain2014,jain2017}. These investigations established an important theoretical framework for predicting the hierarchy of instabilities in electron-scale current sheets and highlighted the role of current-sheet geometry in determining the dominant mode.

While these EMHD studies provide important predictions for the dominant instability regime, they do not fully capture the nonlinear evolution and interaction between competing modes in fully kinetic systems. In particular, it remains unclear how accurately EMHD mode-selection predictions persist in kinetic regimes, and how the interplay between tearing, surface-preserving, and shear-driven instabilities unfolds during the nonlinear evolution of localized current sheets.

Fully kinetic particle-in-cell (PIC) simulations provide a natural framework for investigating electron-scale current sheets in collisionless plasmas. By self-consistently evolving particle dynamics and electromagnetic fields, PIC simulations capture kinetic effects such as particle trapping, pressure tensor contributions, and nonlinear wave–particle interactions that are not included in EMHD models. Previous kinetic studies have confirmed the development of electron inertial tearing modes and demonstrated the formation of electron-scale magnetic islands, current-sheet thinning, and complex nonlinear dynamics \cite{zenitani2005,schoeffler2011,hesse2018}. Multidimensional PIC simulations have further revealed current-sheet filamentation, vortex formation, and coupling between reconnection and shear-driven instabilities in turbulent collisionless plasmas \cite{karimabadi2013,sioulas2022}. However, most previous kinetic studies have focused on either reconnection in extended current layers or turbulent current-sheet formation, rather than the controlled evolution of localized electron-scale current sheets.

Consequently, a systematic understanding of the early and late nonlinear evolution of localized electron current sheets remains limited. In particular, the temporal ordering and mutual interaction of tearing, surface-preserving, and shear-driven instabilities in fully kinetic systems have not been examined in a controlled manner. Although previous studies have shown that secondary Kelvin–Helmholtz–like instabilities can develop within magnetic islands during reconnection \cite{delsarto2005}, the dependence of such processes on current-sheet thickness and dimensionality, as well as their role in determining the dominant nonlinear pathway, remains poorly understood.

In this work, we investigate the stability and nonlinear evolution of localized electron-scale current sheets using fully kinetic particle-in-cell (PIC) simulations in both two and three spatial dimensions. Two equilibrium configurations corresponding to thin and wide current sheets ($\epsilon=0.3$ and $\epsilon=0.9$) are considered, allowing us to examine how current-sheet thickness influences the hierarchy of instabilities. By comparing the simulation results with predictions from linear EMHD theory, we assess the validity of EMHD mode-selection predictions and examine how kinetic effects modify the nonlinear evolution. The simulations reveal a clear thickness-dependent transition in the dominant instability: while thin sheets remain primarily tearing-dominated, wider sheets develop strong shear-driven vortex structures characteristic of a Kelvin–Helmholtz–type instability in three dimensions. 

More importantly, we identify a clear nonlinear sequence of instability development in fully kinetic systems. For wider current sheets, the evolution proceeds through an initial shear-dominated phase, followed by saturation of the Kelvin–Helmholtz–type instability and the subsequent re-emergence of tearing dynamics as the dominant mechanism responsible for magnetic topology change. In contrast, thin current sheets remain tearing-dominated throughout, with no transition to a shear-driven regime. 

These results provide new insight into how dimensionality and current-sheet geometry control not only the dominant instability, but also the temporal ordering of nonlinear processes in electron-scale current sheets, with direct implications for electron-scale reconnection and current-sheet stability in collisionless plasmas..

The remainder of this paper is organized as follows. In Sec.~II, we describe the numerical method, the equilibrium configuration, and the initial and boundary conditions used in the simulations. Section~III presents the results obtained from the two- and three-dimensional simulations. In Sec.~IV, we provide a detailed comparison between the two- and three-dimensional evolution of the system for the two current-sheet thicknesses considered. Finally, Sec.~V summarizes the main conclusions of this work.

\section{\label{sec:level1}Simulation Model and Numerical Setup}
We perform fully kinetic Particle-in-Cell (PIC) simulations using the OSIRIS 4.0 framework to investigate the evolution of electron-scale tearing instabilities in thin current sheets in both two and three dimensions \cite{fonseca2002}. The PIC approach self-consistently evolves particle dynamics and electromagnetic fields by solving the Vlasov–Maxwell system, thereby retaining electron inertia and fully nonlinear electromagnetic interactions at electron scales. This fully kinetic treatment avoids the need for fluid closures and enables a self-consistent description of the evolution of electron-scale current sheets, particularly during the linear growth and nonlinear saturation of the instability.

\subsection{Equilibrium Configuration and Initial Conditions}
\begin{figure}[h] 
\centering 
\includegraphics[width=0.8\linewidth]{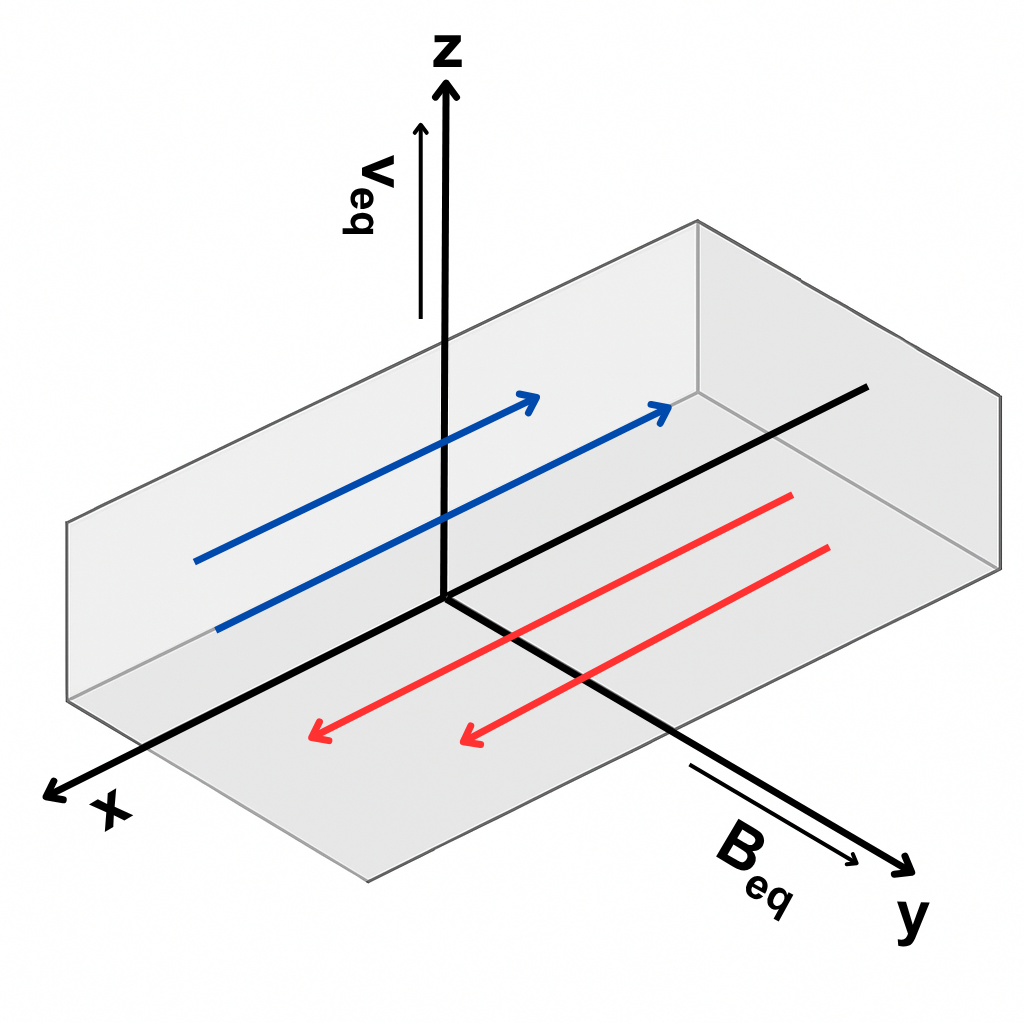} \caption{Schematic of the simulation geometry. The equilibrium magnetic field $\mathbf{B}_{\mathrm{eq}}$ is along $y$ and varies across $x$, forming a localized current sheet. The equilibrium electron flow $\mathbf{v}_{\mathrm{eq}}$ is along $z$. The $x$-direction is the gradient (inflow) direction, and $y$ corresponds to the magnetic field (outflow) direction.} 
\label{fig:setup} 
\end{figure}

The simulations are performed in a three-dimensional Cartesian domain with coordinates $(x,y,z)$, as illustrated in Fig.~\ref{fig:setup}. We consider an electron-only current sheet in which the equilibrium magnetic field is directed along $\hat{y}$ and varies across $\hat{x}$,
\begin{equation}
\mathbf{B}_{\mathrm{eq}}(x) = \hat{y}\left[B_{00}\,\mathrm{sech}(x/\epsilon)\tanh(x/\epsilon)\right],
\label{B}
\end{equation}
where $\epsilon$ characterizes the shear width and $B_{00}$ sets the field amplitude. This profile produces a localized magnetic shear layer centered in the domain, enabling the study of electron-scale instabilities without imposing large-scale reconnection geometry.

Assuming stationary ions ($\mathbf{v}_i = 0$), the equilibrium current is entirely supported by electron motion. From Amp\`ere’s law, the corresponding electron flow is directed along $\hat{z}$,
\begin{equation}
\mathbf{v}_{\mathrm{eq}} = -\nabla\times\mathbf{B}_{\mathrm{eq}} = \hat{z}\,v_0(x)
= \hat{z}\,\frac{B_{00}\,\sinh^2(x/\epsilon)}{\epsilon\,\cosh^3(x/\epsilon)}.
\label{v}
\end{equation}

In this configuration, ions provide a stationary, uniform neutralizing background, isolating the electron-scale dynamics and suppressing ion-timescale effects. The system therefore corresponds to the inertial electron magnetohydrodynamic (EMHD) limit, in which the evolution is governed by electron inertia and current-driven magnetic shear.

Unlike the classical Harris sheet, which relies on pressure balance between magnetic and plasma pressures, the present configuration is compatible with the cold, electron-only framework adopted here, where thermal pressure is absent. The chosen profile thus provides a self-consistent way to model a localized current sheet sustained by electron motion and magnetic structure alone.

\begin{figure}[h]
    \includegraphics[width=1.0\linewidth]{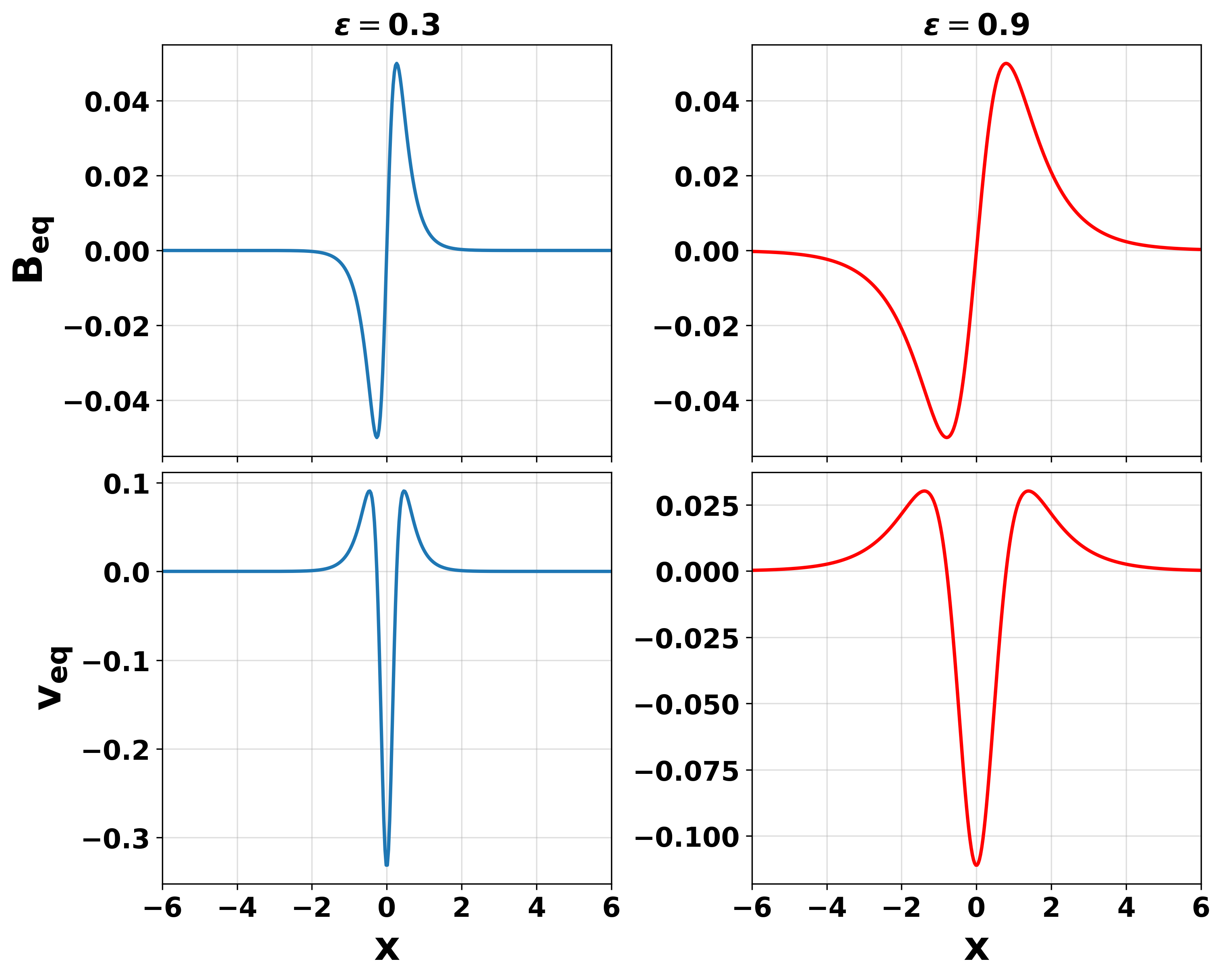}
    \caption{The equilibrium magnetic field (top row) and velocity (bottom row) profiles for two values of the $\epsilon = 0.3$ (left panel) and $\epsilon = 0.9$ (right panel).}
    \label{fig:profile}
\end{figure}

The $x$-direction corresponds to the gradient (inflow) direction, while the magnetic field is aligned along $y$ and the equilibrium current and electron flow are along $z$, as shown in Fig.~\ref{fig:setup}. No explicit perturbations are imposed; instead, the instability develops self-consistently from intrinsic numerical noise, allowing unstable modes to emerge without bias toward a particular wavelength or symmetry. The equilibrium profiles are shown in Fig.~\ref{fig:profile}.

\subsection{Numerical Parameters}

\begin{table}[h]
\centering
\caption{Simulation domain sizes and grid resolutions used in two- and three-dimensional PIC simulations. All lengths are normalized to the electron skin depth $d_e$.}
\label{tab:domain}
\begin{tabular}{c c c c c c}
\hline\hline
Dimension & $\epsilon$ & $L_x/d_e$ & $L_y/d_e$ & $L_z/d_e$ & Grid Resolution \\
\hline
2D & 0.3 & 3.0 & 1.5 & -- & $256 \times 128$ \\
2D & 0.9 & 9.0 & 4.5 & -- & $256 \times 128$ \\
\hline
3D & 0.3 & 3.0 & 1.5 & 1.5 & $256 \times 128 \times 128$ \\
3D & 0.9 & 9.0 & 4.5 & 4.5 & $256 \times 128 \times 128$ \\
\hline\hline
\end{tabular}
\end{table}

Simulations are performed for multiple current-sheet shear widths, $\epsilon = 0.3$ and $\epsilon = 0.9$, in order to investigate the dependence of the growth and saturation of electron-scale tearing instabilities on the thickness of the current layer. All simulations employ the fully relativistic OSIRIS PIC framework, with quantities normalized to characteristic electron-scale parameters. Lengths are normalized to the electron skin depth $d_e = c/\omega_{pe}$, time to the inverse electron plasma frequency $\omega_{pe}^{-1}$, velocities to the speed of light $c$, and magnetic fields to $B_0 = m_e c \omega_{pe}/e$. Unless otherwise stated, all quantities are expressed in these normalized units.

The simulation domain sizes and grid resolutions used in the two- and three-dimensional simulations are summarized in Table~\ref{tab:domain}. The spatial resolution is chosen to adequately resolve both the equilibrium current layer and the linearly unstable electron-scale modes. Two-dimensional (2D) and three-dimensional (3D) simulations are performed with identical physical parameters, differing only in the inclusion of the third spatial dimension.

The two-dimensional simulations employ $10\times5$ particles per cell, while the three-dimensional runs use $10\times5\times5$ particles per cell. Only electrons are treated as kinetic particles, while ions are modeled as a stationary, uniform neutralizing background and are not explicitly evolved in the simulation. Electrons are initialized with a cold (zero-temperature, zero-momentum) distribution, so that the system has no thermal spread at initialization. This cold electron-only setup isolates inertial electron-scale dynamics and enables direct comparison with Electron Magnetohydrodynamics (EMHD) predictions without the influence of thermal-pressure effects. The time step satisfies the Courant condition, $\Delta t \leq 0.99\,\Delta x/c$, and periodic boundary conditions are applied in all spatial directions. The simulation domain is chosen sufficiently large in the $x$ direction such that boundary effects do not influence the evolution of the current sheet or the tearing dynamics.

\subsection{Physical Regimes Considered}

The narrow current-sheet case ($\epsilon = 0.3$) is associated with stronger magnetic shear and larger electron drift velocities ($v \simeq 0.3c$), whereas the wider sheet ($\epsilon = 0.9$) exhibits weaker shear with lower drift velocities ($v \simeq 0.1c$). The corresponding equilibrium magnetic-field and electron flow profiles are shown in Fig.~\ref{fig:profile}.

These two configurations provide a controlled framework for examining how current-sheet thickness and electron flow strength influence the linear growth and nonlinear evolution of electron-scale instabilities in two and three dimensions.

\section{Results}
We now examine the evolution of electron-scale instabilities in the localized current-sheet configuration using fully kinetic PIC simulations. Particular attention is given to the role of current-sheet thickness and dimensionality in determining the dominant instability and subsequent nonlinear dynamics. Simulations are performed for two equilibrium shear widths, $\epsilon=0.9$ (wide sheet) and $\epsilon=0.3$ (thin sheet), in both two and three spatial dimensions. 

We first establish the two-dimensional baseline, where the instability is governed by electron-scale physics alone, and compare the measured linear growth rates with predictions from linear EMHD theory. We then examine how nonlinear evolution depends on current-sheet thickness before turning to the fully three-dimensional dynamics.

\subsection{Linear growth and equilibrium diagnostics}
\label{sec:linear}

A useful diagnostic for identifying regions susceptible to different EMHD instabilities is the quantity $B_0 B_0''$, which distinguishes tearing-favorable and surface-preserving regions. In EMHD, tearing modes arise where $B_0 B_0'' < 0$, allowing changes in magnetic topology, whereas regions satisfying $B_0 B_0'' > 0$ support non-tearing, surface-preserving modes \cite{gaur2016}. Both instabilities originate from current gradients and magnetic-field curvature; however, they differ in their spatial character and nonlinear behavior. Tearing is intrinsically a nonlocal instability that depends on the global structure of the current sheet, whereas the surface-preserving mode is a local instability, typically associated with short-wavelength perturbations satisfying $k\epsilon \gg 1$, and does not involve reconnection.

\begin{figure}[H]
\centering
\includegraphics[width=0.8\linewidth]{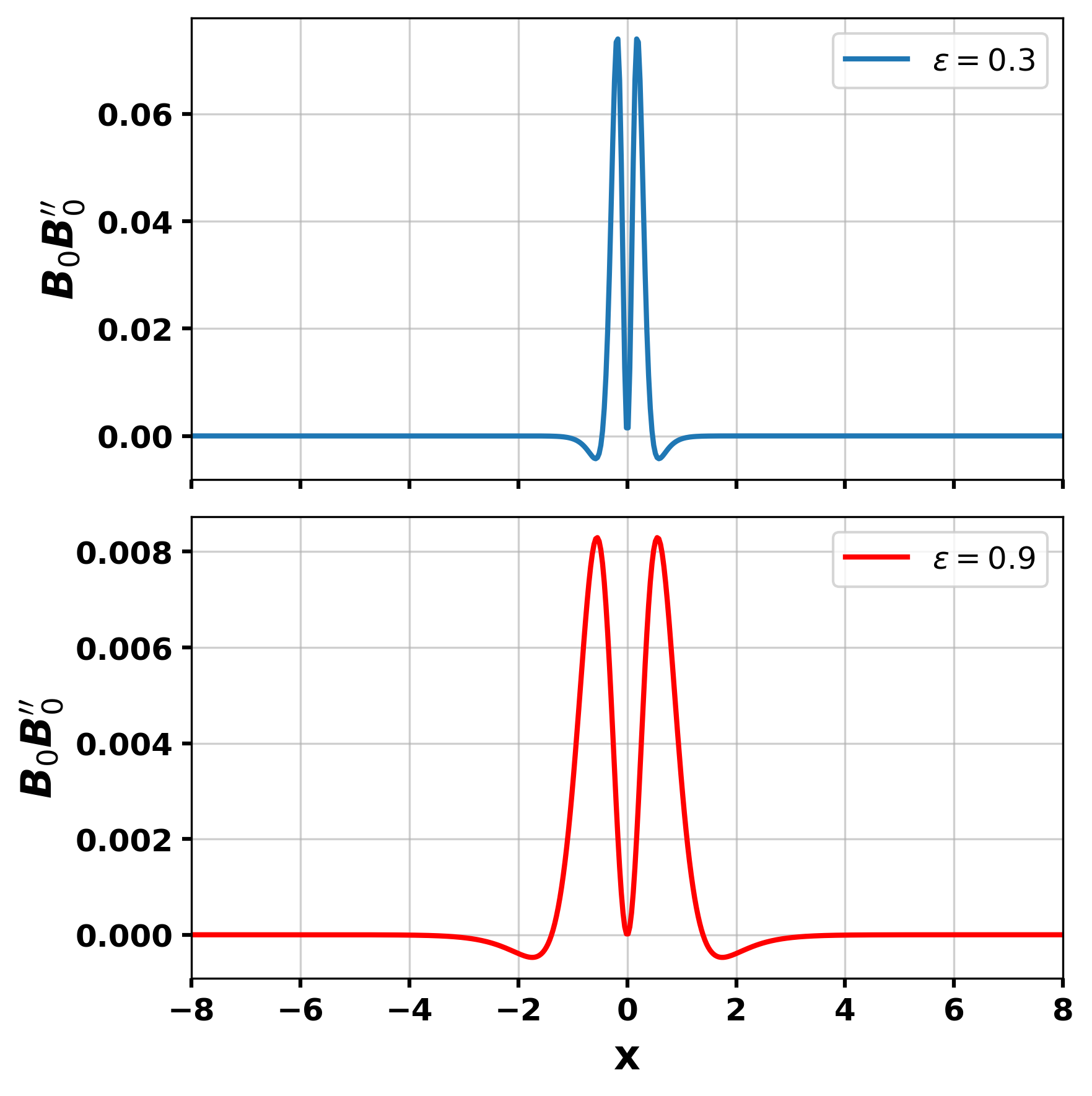}
\caption{Variation of the quantity $B_0 B_0''$ across the current sheet for (top) $\epsilon=0.3$ and (bottom) $\epsilon=0.9$. Regions with $B_0 B_0''<0$ correspond to tearing-favorable locations, while $B_0 B_0''>0$ identifies regions where surface-preserving EMHD modes may develop. The stronger curvature for $\epsilon=0.3$ indicates enhanced susceptibility to non-tearing dynamics.}
\label{bdv}
\end{figure}

Figure~\ref{bdv} shows the spatial variation of $B_0 B_0''$ across the current sheet for both equilibrium widths considered. For the thinner sheet ($\epsilon=0.3$), the magnitude of $B_0 B_0''$ is significantly larger and exhibits pronounced regions of both signs near the center of the sheet. In contrast, for the wider sheet ($\epsilon=0.9$), the curvature is weaker and the regions favorable to non-tearing activity are much less prominent. This difference anticipates a stronger influence of shear-driven, non-tearing dynamics in the thin-sheet case.

The linear growth of the instability is quantified using the perturbed total energy ($E_{pert}$), which measures the energy contained in magnetic and velocity fluctuations. Since the perturbed energy scales with the square of field amplitudes, the linear phase is expected to follow
\begin{equation}
E_{\mathrm{pert}} \propto e^{2\gamma t},
\end{equation}
where $\gamma$ is the linear growth rate.

\begin{figure}[h]
\centering
\includegraphics[width=1.0\linewidth]{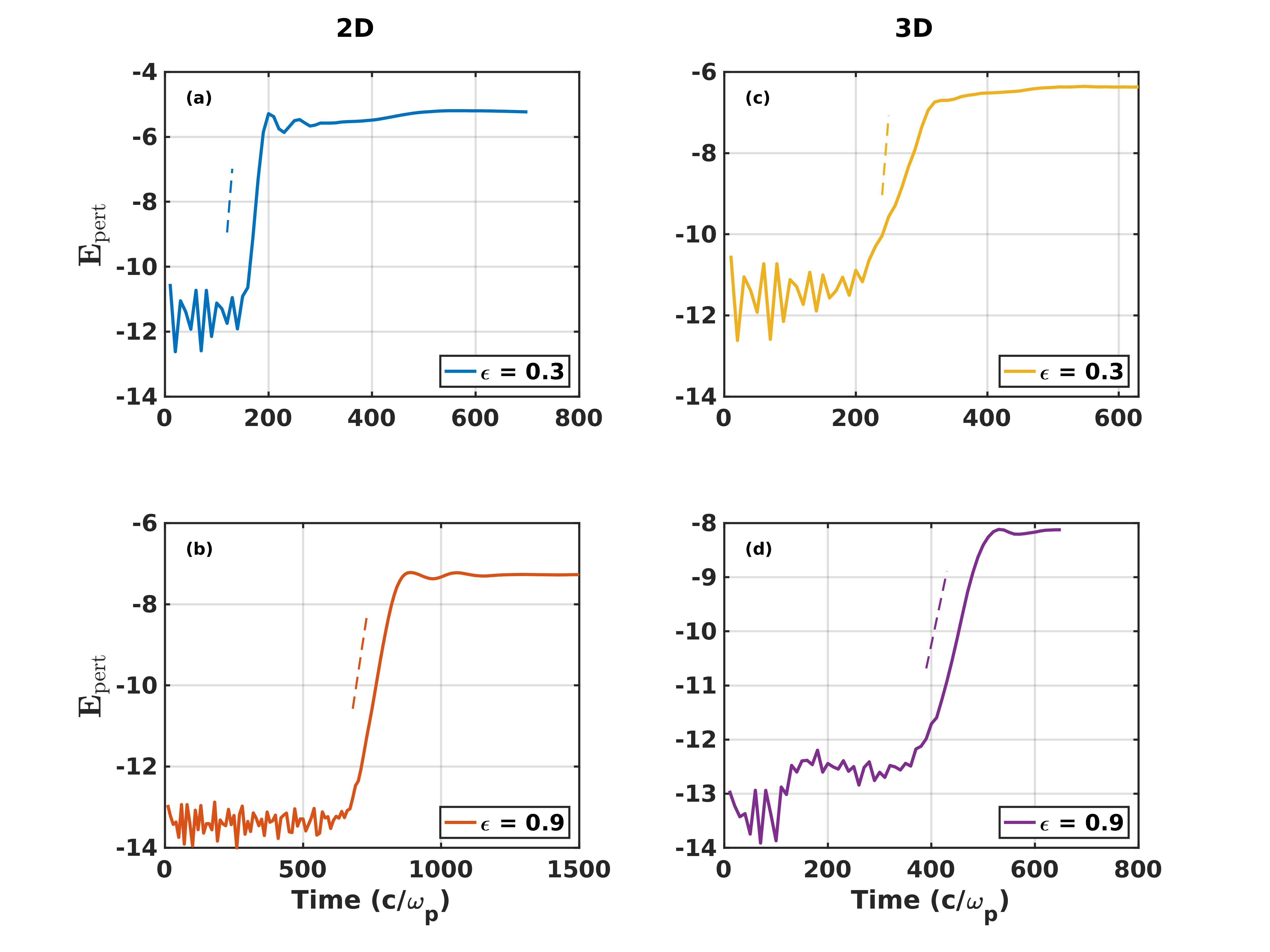}
\caption{Temporal evolution of the logarithm of perturbed total energy $E_{\mathrm{pert}}$ for $\epsilon=0.3$ (top panels) and $\epsilon=0.9$ (bottom panels) in two-dimensional (left) and three-dimensional (right) simulations. Dashed lines denote slopes corresponding to linear EMHD growth rates. The 2D results exhibit exponential growth consistent with electron inertial tearing in both cases. In contrast, the 3D evolution shows a thickness-dependent modification: the thin sheet remains tearing-dominated with reduced effective growth, while the wide sheet exhibits faster growth associated with the onset of shear-driven (Kelvin–Helmholtz) dynamics.}
\label{pte_2d}
\end{figure}

Figure~\ref{pte_2d} shows the temporal evolution of the perturbed total energy ($E_{pert}$) for both values of $\epsilon$ in two- and three-dimensional simulations. In two dimensions (left column), the early-time evolution follows the expected exponential scaling. For the wider current sheet ($\epsilon=0.9$), the theoretical growth rate $\gamma=0.0224$ yields a slope $2\gamma$ that closely matches the exponential growth observed in the PIC simulations. Similarly, for the thinner sheet ($\epsilon=0.3$), the theoretical prediction $\gamma=0.098$ reproduces the linear regime with good accuracy, giving a measured growth rate $\gamma_{\mathrm{PIC}}\simeq0.196$. These results confirm that the linear phase in 2D is governed by electron inertial tearing and is quantitatively consistent with EMHD theory.

In three dimensions (right column), the early-time evolution also exhibits an approximately exponential growth phase; however, the growth rates and subsequent evolution differ from their 2D counterparts. The dashed reference lines in the 3D panels correspond to linear growth rates of the dominant two-dimensional modes: for $\epsilon=0.3$, the slope corresponds to the tearing mode with $k_z=0$, while for $\epsilon=0.9$, it corresponds to the Kelvin–Helmholtz mode with $k_y=0$.

For the thin current sheet ($\epsilon=0.3$), the growth remains broadly consistent with tearing-dominated dynamics; however, the effective growth rate is reduced relative to the 2D tearing prediction, indicating the influence of three-dimensional effects (see Sec.~III.C). In contrast, for the wider sheet ($\epsilon=0.9$), the growth deviates from the 2D tearing behavior and instead reflects the onset of a shear-driven Kelvin–Helmholtz–type instability. In this case, the linear growth rate agrees well with the corresponding two-dimensional Kelvin–Helmholtz prediction. 

Deviations from the linear slope at later times in both dimensions mark the transition to nonlinear evolution, with the nature of the nonlinear state determined by the competition between tearing and shear-driven dynamics in three dimensions.

\subsection{Nonlinear evolution in two dimensions}
\label{sec:nonlinear2d}

\begin{figure}[h] 
\centering 
\includegraphics[width=1.0\linewidth]{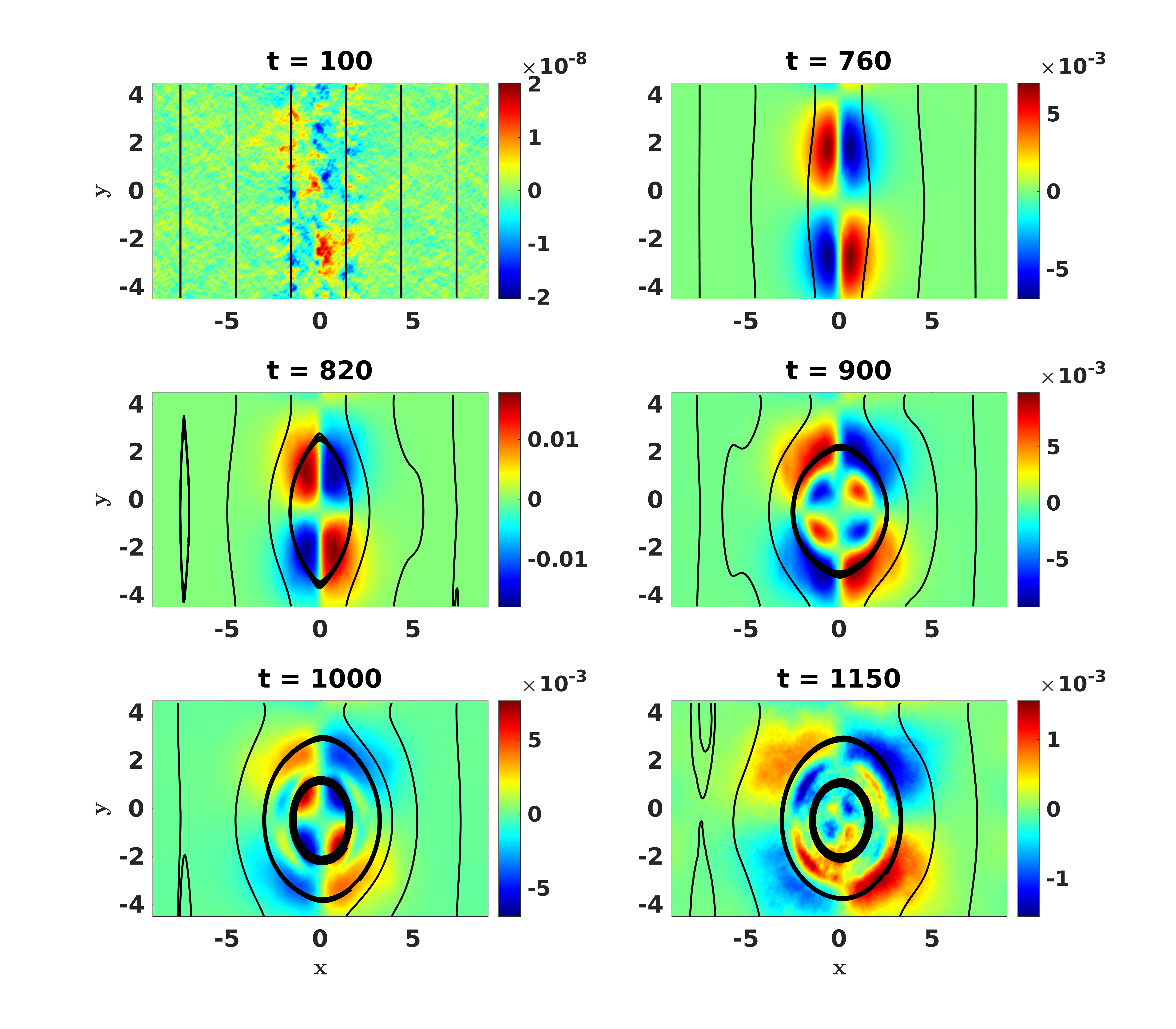} 
\caption{Evolution of the out-of-plane magnetic field $B_z$ in the 2D simulation for $\epsilon=0.9$ (wide sheet), with in-plane magnetic field lines superimposed.} 
\label{bz_2d_wide} 
\end{figure} 

\begin{figure}[h] 
\centering 
\includegraphics[width=1.0\linewidth]{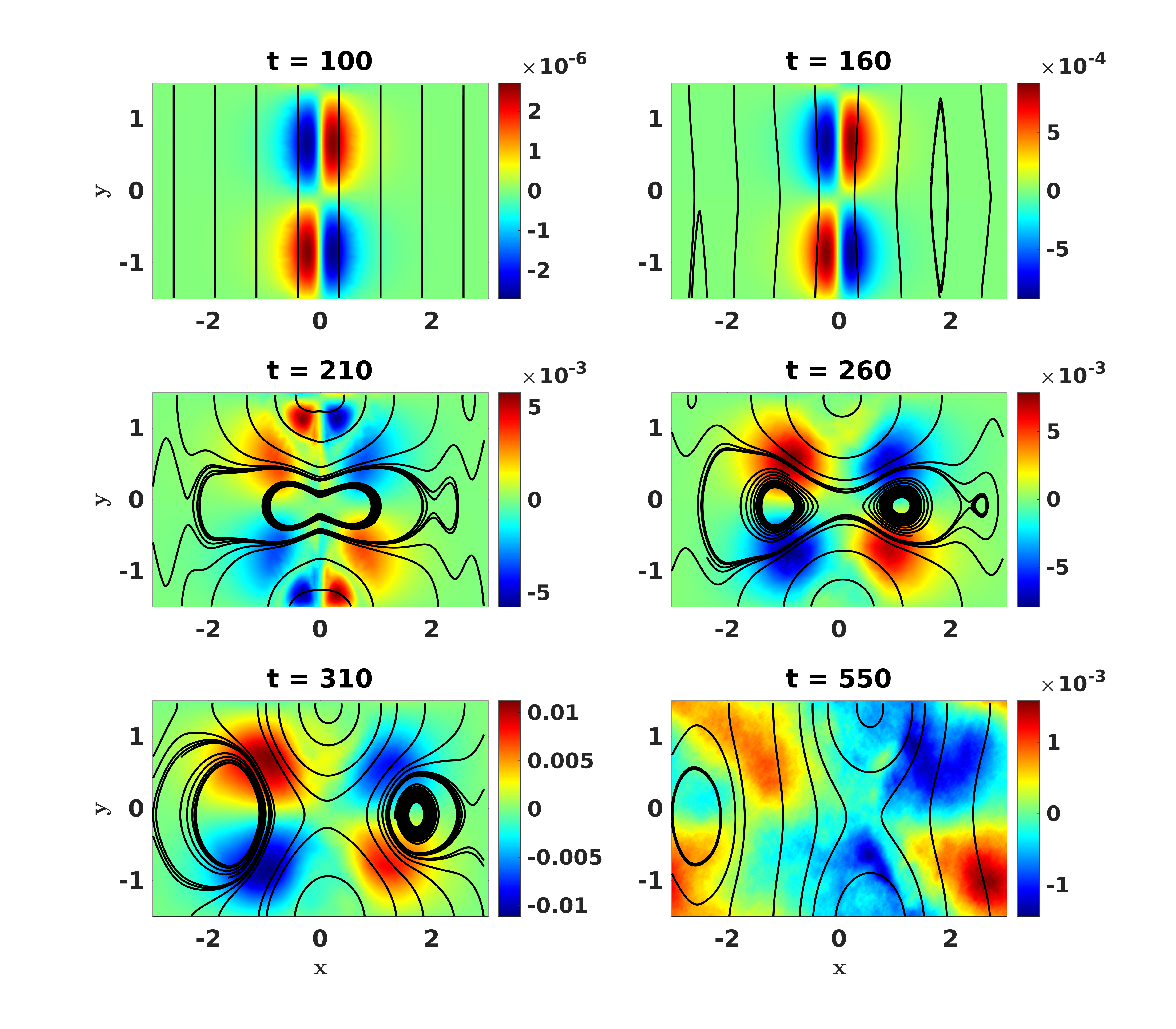} 
\caption{Evolution of the out-of-plane magnetic field $B_z$ in the 2D simulation for $\epsilon=0.3$ (thin sheet), with in-plane magnetic field lines superimposed.} 
\label{bz_2d_thin} 
\end{figure}

We now examine the evolution of the system in two-dimensional simulations, focusing on the development and nonlinear interaction of magnetic structures. The dynamics are visualized using contour plots of the out-of-plane magnetic field $B_z$, with in-plane magnetic field lines superimposed (Figs.~\ref{bz_2d_wide} and~\ref{bz_2d_thin}).

At early times, magnetic field lines remain straight and follow the equilibrium configuration, while $B_z$ exhibits small-amplitude fluctuations arising from intrinsic numerical noise. As the system enters the linear growth phase, these fluctuations organize into coherent structures, and a characteristic quadrupolar $B_z$ pattern develops around the neutral line, indicating the onset of reconnection associated with electron inertial tearing.

The subsequent nonlinear evolution depends strongly on the current-sheet thickness. For the wide sheet ($\epsilon = 0.9$), the small magnitude of $B_0 B_0''$ near the center implies that tearing governs the evolution. As shown in Fig.~\ref{bz_2d_wide}, a single coherent magnetic island forms at the neutral line and remains symmetric about $x = 0$. The nonlinear dynamics closely resemble classical EMHD tearing, with steady island growth and smooth deformation of the current layer.

In contrast, the thin-sheet case ($\epsilon = 0.3$) exhibits qualitatively different behavior. The equilibrium contains regions with $B_0 B_0'' > 0$ adjacent to the neutral line (Fig.~\ref{bdv}), which support surface-preserving EMHD modes in addition to tearing. As a result, the instability involves the coexistence of global tearing and localized surface-preserving perturbations. Linear EMHD analyses \cite{gaur2016} show that while the tearing mode is symmetric about the neutral line, the surface-preserving mode is intrinsically asymmetric and localized away from the center. In the thin-sheet regime, the superposition of these modes therefore introduces an inherent asymmetry in the unstable eigenstructure.

This is reflected in Fig.~\ref{bz_2d_thin}, where the primary island initially forms at the neutral line but exhibits a clear structural asymmetry from early times. This asymmetry leads to an imbalance in magnetic tension and electron flow across the island, causing it to drift away from the neutral line as the system evolves. At later times, the island settles at a finite offset from $x=0$, accompanied by a partial reduction of its asymmetry, indicating relaxation toward a more dynamically balanced configuration.

The asymmetry is further evident in the secondary islands that form on either side of the neutral line, which are not identical in strength or structure. These islands arise self-consistently from the presence of multiple unstable regions associated with strong curvature gradients and the coexistence of tearing and surface-preserving modes. Seconday islands are weaker for $\epsilon = 0.9$ case. As the system evolves, the secondary islands convect along the current sheet under the influence of electron flows, interact with the primary island, and undergo merging or collision, resulting in a dynamically evolving multi-island state.

In the late nonlinear stage, the two configurations follow distinct saturation pathways. For the wider current sheet ($\epsilon = 0.9$), magnetic energy remains concentrated within a single dominant island. The strong bidirectional electron outflows generate shear layers along the separatrices and within the island interior. These shear layers become unstable to a secondary Kelvin--Helmholtz type instability in the EMHD regime, leading to vortical distortions and the onset of small-scale turbulence \cite{delsarto2005}. As these fluctuations intensify, the initially coherent quadrupolar $B_z$ structure becomes progressively distorted, marking a transition from organized tearing to shear-driven turbulent mixing.

In contrast, the thin-sheet case evolves toward a multi-island state governed by the coupled action of tearing and surface-preserving modes. Because magnetic energy is distributed among multiple interacting structures, no single coherent shear layer develops, and the reconnection outflows remain fragmented and time-dependent. This suppresses the onset of secondary Kelvin--Helmholtz type instability within the primary island.

As the evolution proceeds, the secondary islands undergo coalescence, forming larger structures that continue to interact with the primary island. This hierarchical merging leads to continuous redistribution of magnetic flux and current within the system.

At later times, the initially asymmetric configuration partially relaxes through this coalescence process, leading to a more symmetric large-scale structure, although signatures of multi-island interaction persist. The quadrupolar $B_z$ pattern remains identifiable, and saturation proceeds primarily through island merging rather than shear-driven turbulence.

These results establish a clear two-dimensional baseline. Wide current sheets evolve through symmetric EMHD tearing, followed by the development of shear layers that drive Kelvin--Helmholtz instability and lead to turbulent saturation. In contrast, thin sheets exhibit intrinsically asymmetric evolution due to the coexistence of tearing and surface-preserving modes, resulting in island displacement, multi-island interaction, and coalescence-driven relaxation.

This thickness-dependent distinction highlights the competing roles of magnetic curvature and electron shear in determining the nonlinear pathway. The contrast between these regimes becomes even more pronounced when three-dimensional effects are introduced, as discussed in the following subsection.

\subsection{Nonlinear evolution in three dimensions}
\label{sec:3d}
\begin{figure}[h]
    \centering
    \includegraphics[width=1.0\linewidth]{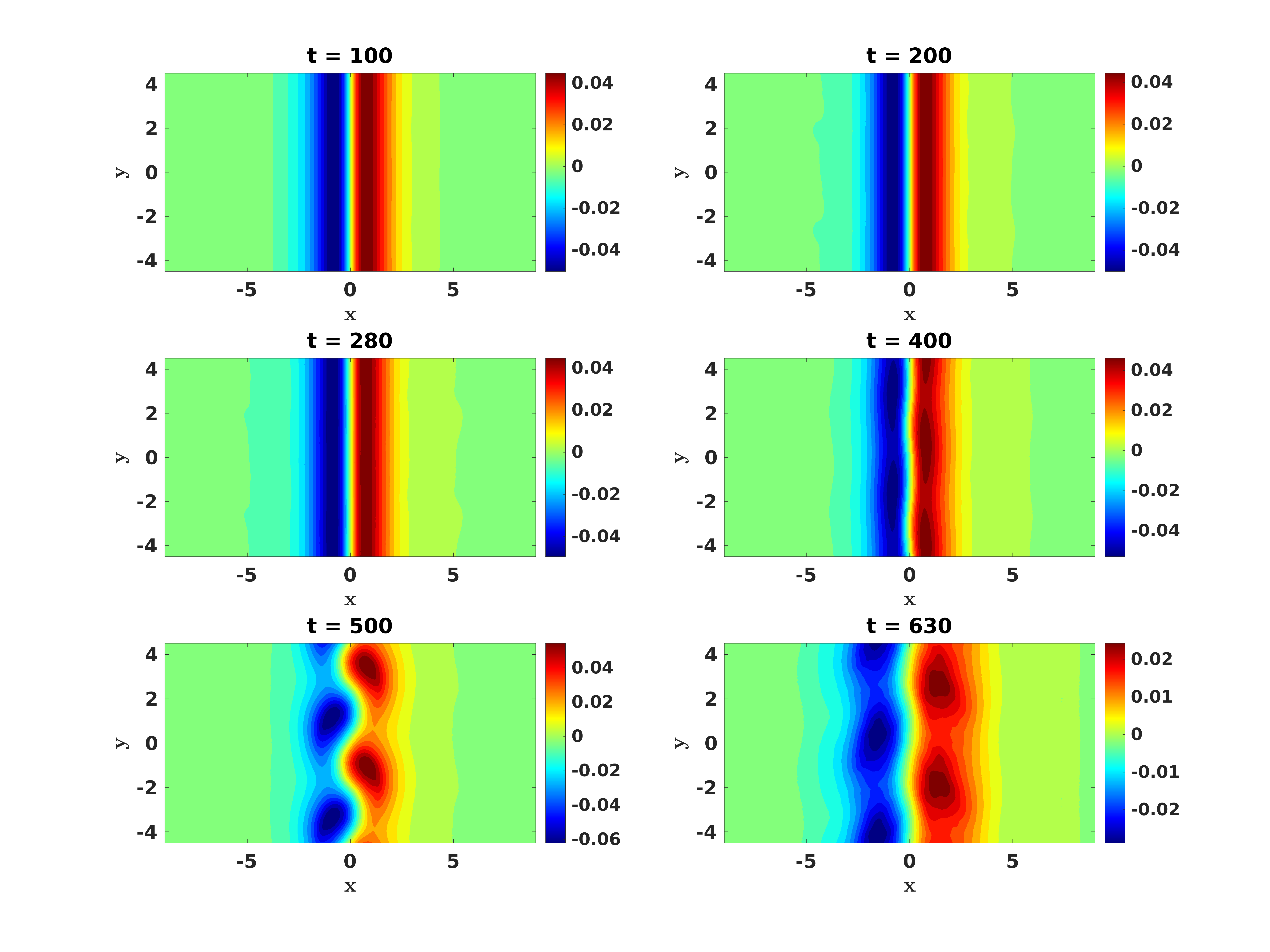}
    \caption{Evolution of the magnetic field component $B_y$ in the $xz$ plane ($y=0$) for the 3D simulation with $\epsilon=0.9$.}
    \label{by_xz_eps0p9}
\end{figure}

\begin{figure}[h]
    \centering
    \includegraphics[width=1.0\linewidth]{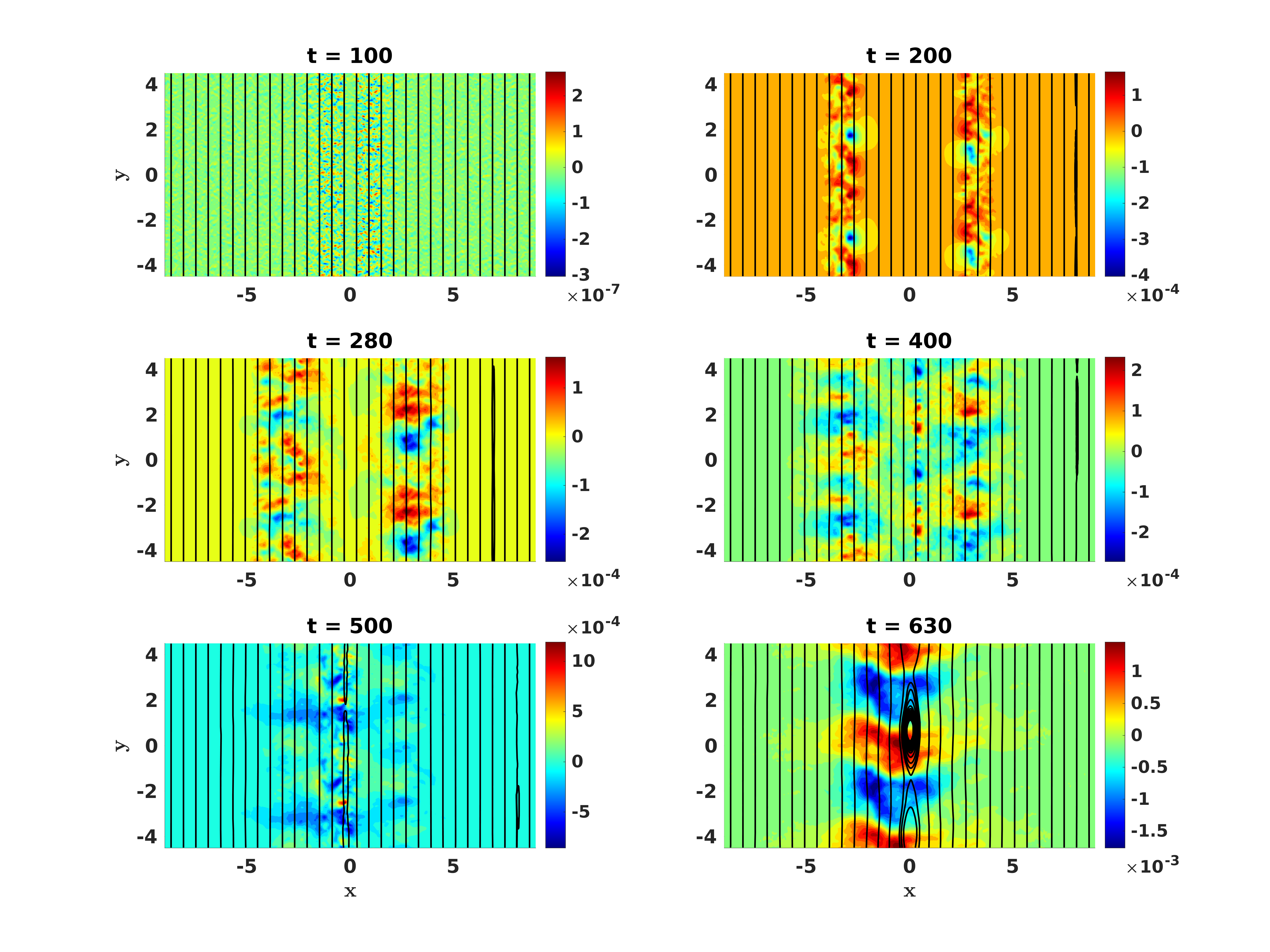}
    \caption{Evolution of the out-of-plane magnetic field $B_z$ in the $xy$ plane ($z=0$) for $\epsilon=0.9$ in the 3D simulation, with magnetic field lines superimposed.}
    \label{bz_xy_eps0p9}
\end{figure}

We now examine the fully three-dimensional evolution of the system, where perturbations along the equilibrium flow direction are allowed. In contrast to the two-dimensional case, where $\partial/\partial z = 0$ suppresses shear-driven instabilities. The inclusion of the third dimension enables velocity-shear–driven modes to develop and interact with tearing dynamics. The evolution is analyzed using $B_z$ in the $xy$ plane, which captures reconnection and magnetic topology, and $B_y$ in the $xz$ plane, which reveals modulation along the current direction and serves as a diagnostic for shear-driven dynamics.

\subsubsection*{Wide sheet: $\epsilon=0.9$}

For the wider current sheet, the three-dimensional evolution differs qualitatively from the two-dimensional case. The $xz$ plane evolution of $B_y$ (representing the flow lines in the $xz$ plane), shown in Fig.~\ref{by_xz_eps0p9}, exhibits the development of clear alternating structures along the current direction, which later evolve into vortex-like patterns. These features are characteristic of a Kelvin–Helmholtz-type instability driven by the equilibrium electron velocity shear.

The corresponding $xy$ plane evolution of $B_z$ (Fig.~\ref{bz_xy_eps0p9}) indicates that no tearing signatures are present at early times; however, a magnetic island appears later, which is elongated and distorted. The reconnection structures are strongly modulated along the third dimension, leading to thin, stretched island-like structures rather than the coherent, symmetric islands observed in two dimensions at early times.

The measured growth rate agrees well with linear theory predictions for the dominant three-dimensional shear mode, confirming that the Kelvin–Helmholtz–type instability governs the early and intermediate nonlinear evolution for $\epsilon=0.9$. Tearing is therefore present but becomes secondary, with its structures continuously reshaped by shear-driven dynamics.

The appearance of a magnetic island in the $xy$ plane at later times ($t \gtrsim 600$), although elongated, indicates that tearing is not suppressed but rather temporarily masked by the faster-growing shear-driven instability during the early and intermediate stages. As the Kelvin–Helmholtz–type mode saturates and redistributes the velocity shear, tearing re-emerges as the dominant mechanism driving magnetic topology change. The resulting island thus reflects tearing evolution on a background modified by prior shear-driven dynamics.

The nonlinear evolution thus exhibits a sequential interplay between shear-driven and reconnection-driven processes.

\begin{figure}[h]
    \centering
    \includegraphics[width=1.0\linewidth]{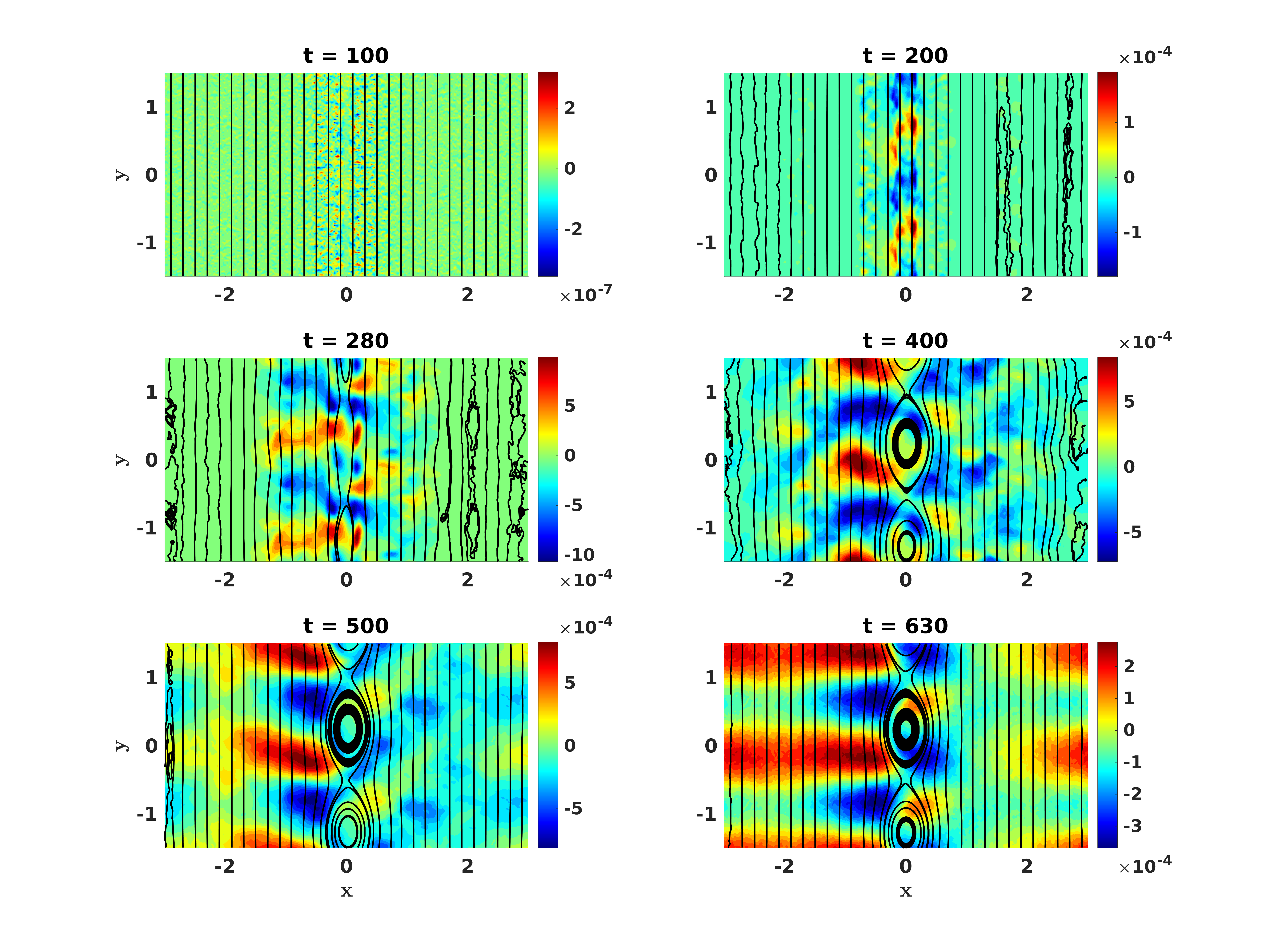}
    \caption{Evolution of the out-of-plane magnetic field $B_z$ in the $xy$ plane ($z=0$) for $\epsilon=0.3$ in the 3D simulation, with magnetic field lines superimposed.}
    \label{bz_xy_eps0p3}
\end{figure}

\begin{figure}[h]
    \centering
    \includegraphics[width=1.0\linewidth]{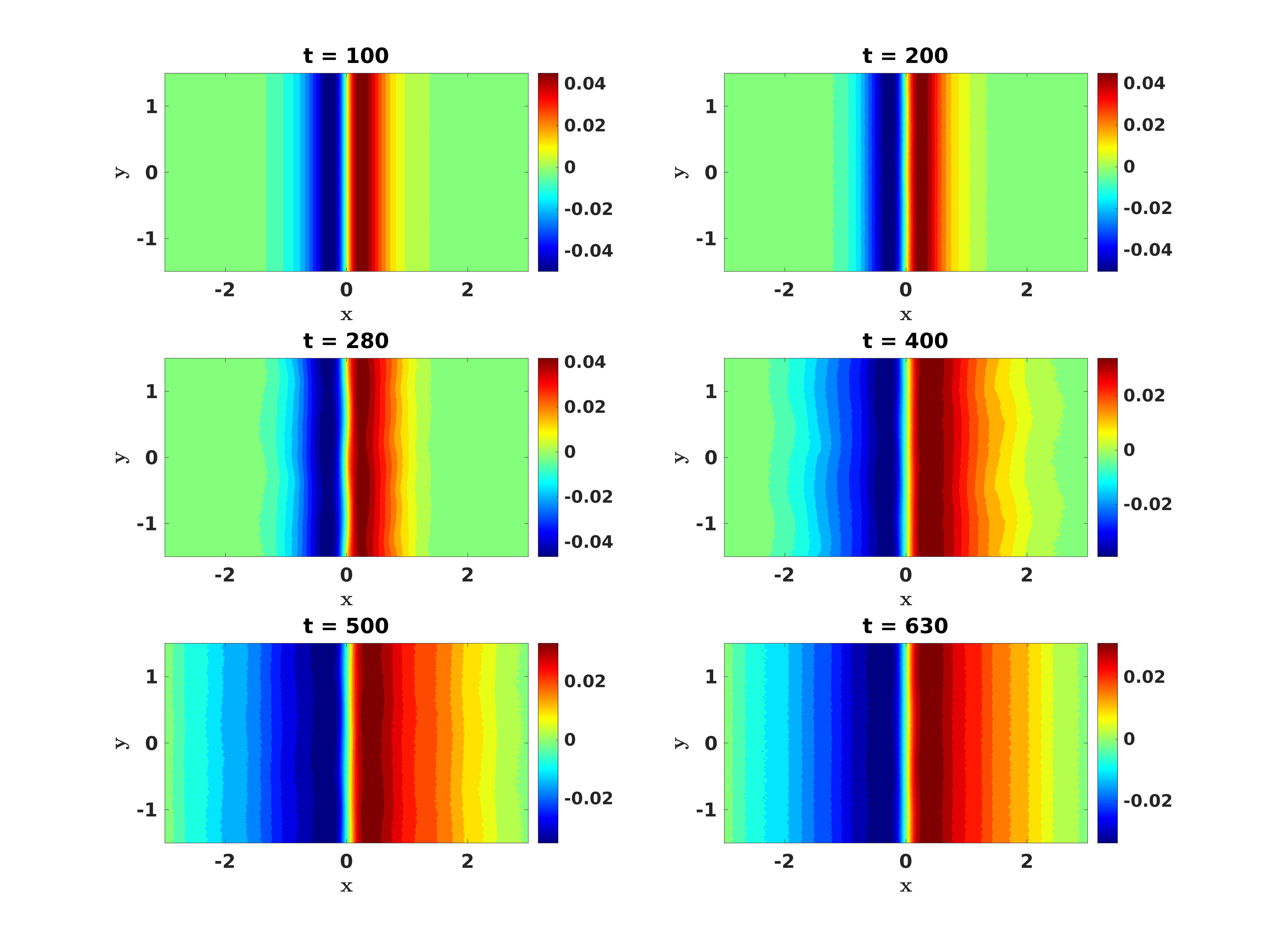}
    \caption{Evolution of the magnetic field component $B_y$ in the $xz$ plane ($y=0$) for the 3D simulation with $\epsilon=0.3$.}
    \label{by_xz_eps0p3}
\end{figure}

\subsubsection*{Thin sheet: $\epsilon=0.3$}

In contrast, the thin current sheet exhibits a qualitatively different three-dimensional evolution. The $xy$ plane evolution of $B_z$ (Fig.~\ref{bz_xy_eps0p3}) shows the rapid formation of multiple magnetic islands with the characteristic quadrupolar structure associated with electron inertial tearing. These islands remain relatively coherent and are distributed along the current sheet, closely resembling the structures observed in the two-dimensional simulations.

A slight asymmetry in the island structure is observed (at $t = 280$,) consistent with the influence of surface-preserving perturbations identified in the two-dimensional simulations. In contrast to the two-dimensional case, secondary islands do not interact with the primary island, and the primary island does not exhibit gradual drift away from the nutral line.

The corresponding $xz$ plane evolution of $B_y$ (Fig.~\ref{by_xz_eps0p3}) does not exhibit the clear vortex roll-up or alternating patterns characteristic of a Kelvin–Helmholtz instability. Instead, only weak distortions and localized modulation of the flow-aligned structures are observed. These distortions do not develop into coherent vortices, indicating that shear-driven dynamics remain subdominant and do not manifest as a fully developed Kelvin–Helmholtz–type instability in this case.

This behavior is consistent with linear theory, which predicts tearing to be the fastest-growing mode for $\epsilon = 0.3$. However, the growth rate measured in fully kinetic three-dimensional simulations is lower than the theoretical prediction. This reduction suggests that, although tearing remains the dominant instability, its effective growth is moderated by three-dimensional effects, including mode coupling and the redistribution of energy among multiple modes.

The discrepancy between the measured growth rates and their two-dimensional counterparts can be attributed to the inherently multimode nature of the three-dimensional system, in which energy is shared among competing instabilities and finite-$k_z$ perturbations modify the effective linear growth. A comprehensive three-dimensional linear stability analysis is required to accurately determine the growth rate and eigenstructure of the fastest-growing mode; however, such an analysis lies beyond the scope of the present work.

Consistently, no transition to a shear-dominated regime is observed for $\epsilon=0.3$, even in fully three-dimensional simulations.

\subsubsection*{Mode competition and dimensional transition}

The comparison between the two cases demonstrates that the dominant instability in three dimensions is controlled by the competition between curvature-driven and shear-driven dynamics. For $\epsilon=0.3$, strong magnetic curvature promotes tearing and surface-preserving modes, suppressing the development of Kelvin–Helmholtz instability and resulting in similar behavior in two and three dimensions. For $\epsilon=0.9$, weaker curvature allows velocity shear to dominate, leading to the emergence of Kelvin–Helmholtz–type instability and a qualitatively different nonlinear evolution in three dimensions.

These results establish a clear thickness-dependent transition: thin current sheets are curvature-dominated and remain tearing-controlled even in three dimensions, whereas wider current sheets are shear-dominated and develop Kelvin–Helmholtz–type dynamics when three-dimensional perturbations are allowed. This highlights the critical role of dimensionality in enabling shear-driven instabilities and demonstrates that two-dimensional models are sufficient for thin sheets but insufficient for capturing the dynamics of wider current layers.

\section{Comparison between two- and three-dimensional dynamics}
\label{sec:comparison}

The results presented above enable a direct comparison between the two- and three-dimensional evolution of electron-scale instabilities in localized current sheets. This comparison highlights the critical role of dimensionality in determining both the dominant instability and the nonlinear pathways through which magnetic energy is redistributed.

In two dimensions, the evolution is governed primarily by electron inertial tearing. For both current-sheet thicknesses considered, the linear growth rates measured in the simulations agree closely with predictions from linear EMHD theory (Fig.~\ref{pte_2d}), confirming that the early-time dynamics are well described by electron-scale tearing physics. The nonlinear evolution proceeds through the formation and growth of magnetic islands accompanied by the characteristic quadrupolar $B_z$ structure (Figs.~\ref{bz_2d_wide} and \ref{bz_2d_thin}). While the wide sheet ($\epsilon=0.9$) exhibits symmetric island growth centered at the neutral plane, the thin sheet ($\epsilon=0.3$) shows weak asymmetry and gradual island displacement. As demonstrated by the curvature profile in Fig.~\ref{bdv}, this asymmetry arises from the presence of regions where $B_0 B_0''>0$, allowing a surface-preserving EMHD mode to coexist with tearing. Nevertheless, in two dimensions, tearing remains the dominant mechanism responsible for magnetic topology change in both cases.

The inclusion of the third spatial dimension introduces a qualitatively new degree of freedom by allowing perturbations along the equilibrium flow direction. This enables velocity-shear–driven instabilities to develop and compete with tearing, fundamentally altering the nonlinear dynamics. The relative importance of these instabilities is found to depend sensitively on current-sheet thickness.

For the wide current sheet ($\epsilon=0.9$), the three-dimensional evolution is initially dominated by a Kelvin–Helmholtz–type instability driven by electron velocity shear. This results in vortex formation, strong modulation of the current layer, and significant distortion of reconnecting structures. However, this shear-dominated phase is transient: following the saturation of the Kelvin–Helmholtz–type instability, tearing re-emerges at later times and leads to the formation of a elongated but coherent magnetic island. This demonstrates that, although shear-driven dynamics can dominate the early nonlinear evolution, magnetic reconnection remains essential in determining the late-time behavior.

For the thin current sheet ($\epsilon=0.3$), the dynamics remain primarily tearing-dominated even in three dimensions. The evolution of magnetic islands closely resembles the two-dimensional case, with only moderate three-dimensional modulation. No coherent Kelvin–Helmholtz vortices develop, indicating that shear-driven instabilities do not become dominant. Instead, the nonlinear evolution reflects the coexistence of tearing and surface-preserving dynamics, leading to asymmetric island growth and enhanced structural complexity. The reduction in growth rate relative to linear theory can be attributed to three-dimensional mode coupling and redistribution of energy among interacting modes, rather than a transition to a shear-driven regime. However, in contrast to the two-dimensional case, secondary islands remain very weak, and the primary island does not exhibit the gradual drift observed in two dimensions.

These results demonstrate that dimensionality does not affect all current sheets uniformly. Rather, it introduces a thickness-dependent transition in the dominant instability mechanism. Thin current sheets remain curvature-dominated and tearing-controlled even in three dimensions, whereas wider current sheets become susceptible to shear-driven instabilities that significantly modify the nonlinear evolution.

A key outcome of this study is the identification of a nonlinear sequence of instability development in three dimensions. For wider current sheets, the evolution proceeds through an initial shear-dominated phase, followed by saturation of the Kelvin–Helmholtz–type instability and subsequent re-emergence of tearing. In contrast, thin current sheets do not exhibit such a transition and remain tearing-dominated throughout. This highlights the role of current-sheet geometry in controlling not only the dominant instability, but also the temporal ordering of nonlinear processes.

Overall, these findings demonstrate that while two-dimensional models capture the essential features of electron-scale tearing, fully three-dimensional simulations are required to identify the dominant instability regime and to accurately describe the nonlinear evolution of electron-scale current sheets, particularly in regimes where velocity shear plays a significant role.

\section{Summary}
\label{sec:summary}

In this work, we have investigated the evolution of localized electron-scale current sheets using fully kinetic, electromagnetic particle-in-cell (PIC) simulations in both two and three spatial dimensions. By considering two equilibrium configurations characterized by different shear widths, $\epsilon=0.3$ (thin sheet) and $\epsilon=0.9$ (wide sheet), we systematically examined how current-sheet thickness and dimensionality determine the dominant instability and nonlinear evolution.

In two dimensions, both configurations are governed primarily by electron inertial tearing. The measured linear growth rates agree closely with predictions from linear EMHD theory, confirming that the early-time evolution is well described by electron-scale tearing dynamics. In the nonlinear phase, the wide sheet exhibits symmetric magnetic island formation centered at the neutral plane, while the thin sheet develops weak asymmetry due to the presence of regions with $B_0 B_0''>0$, which allow a surface-preserving EMHD mode to coexist with tearing. Nevertheless, in 2D, tearing remains the dominant mechanism responsible for magnetic topology change in both cases.

The inclusion of the third spatial dimension introduces qualitatively new behavior by enabling velocity-shear–driven instabilities. The three-dimensional simulations reveal a thickness-dependent transition in the dominant instability. For the wide current sheet ($\epsilon=0.9$), a Kelvin–Helmholtz–type instability driven by electron velocity shear dominates the early and intermediate nonlinear evolution, leading to strong modulation and fragmentation of the current layer. However, this shear-dominated phase is transient: following its saturation, tearing re-emerges at later times and produces distorted but coherent magnetic islands. In contrast, for the thin sheet ($\epsilon=0.3$), electron inertial tearing remains the primary instability even in three dimensions, with no transition to a shear-dominated regime, although its effective growth rate is reduced due to mode coupling and nonlinear interactions.

These results demonstrate that current-sheet thickness and dimensionality together determine the hierarchy of electron-scale instabilities. While two-dimensional models capture the essential tearing physics, they can misidentify the dominant instability regime when velocity shear is significant. Fully three-dimensional modeling is therefore required to accurately describe both the dominant instability and the nonlinear evolution of electron-scale current sheets.

A key outcome of this study is the identification of a nonlinear sequence of instability development in three dimensions. In wider current sheets, the system evolves through an initial shear-dominated phase followed by the re-emergence of tearing after saturation of the Kelvin–Helmholtz–type instability, whereas thin sheets remain tearing-dominated throughout. This highlights the role of current-sheet geometry in controlling not only the dominant instability, but also the temporal ordering of nonlinear processes.

The thickness-dependent transition between tearing-dominated and shear-driven regimes identified here has important implications for electron-scale reconnection in laboratory plasmas, collisionless space environments, and high-energy astrophysical systems. In particular, the balance between magnetic curvature and electron shear may determine whether current sheets evolve through coherent reconnection structures or through shear-driven fragmentation and turbulence at electron scales.

\begin{acknowledgments}
The authors acknowledge the use of computational resources provided by the High Performance Computing (HPC) facility at the Inter-University Centre for Astronomy and Astrophysics (IUCAA), Pune, India. We also thank the IUCAA HPC support team for their valuable assistance. The authors are grateful to Dr. Bhavesh G. Patel, Institute for Plasma Research, Gandhinagar, India, for valuable discussions on PIC simulations.
\end{acknowledgments}

\nocite{*}

\input{manuscript.bbl}
\end{document}

%% file: manuscript.bbl
\providecommand{\noopsort}[1]{}\providecommand{\singleletter}[1]{#1}%